\documentclass[aps,pra,reprint,superscriptaddress]{revtex4-1}

\usepackage{amsmath}
\usepackage{amssymb}
\usepackage{hyperref}
\usepackage{graphicx}
\usepackage{epstopdf}
\usepackage{dcolumn}
\usepackage{bm}
\usepackage{xspace}
\usepackage{verbatim}
\usepackage{color}
\newcommand{\bra}[1]{\left\langle #1 \right|}
\newcommand{\ket}[1]{\left| #1 \right\rangle}

\newcommand{\up}{|\uparrow\rangle}
\newcommand{\updag}{\langle\uparrow|}
\newcommand{\down}{|\downarrow\rangle}
\newcommand{\downdag}{\langle\downarrow|}

\hypersetup{colorlinks=true,linkcolor=blue,citecolor=blue, filecolor=blue,urlcolor=blue,breaklinks=true}

\begin{document}
\title{Mechanical qubit-light entanglers in hybrid nonlinear qubit-optomechanics}

\author{Victor Montenegro}
\email{vmontenegro@uestc.edu.cn}
\affiliation{Institute of Fundamental and Frontier Sciences, University of Electronic Science and Technology of China, Chengdu 610051, China}
\affiliation{Department of Physics and Astronomy, University College London, Gower Street, London WC1E 6BT, United Kingdom}

\author{G. D. de Moraes Neto}
\affiliation{Institute of Fundamental and Frontier Sciences, University of Electronic Science and Technology of China, Chengdu 610051, China}
\affiliation{Department of Physics and Astronomy, University College London, Gower Street, London WC1E 6BT, United Kingdom}

\author{Sougato Bose}
\affiliation{Department of Physics and Astronomy, University College London, Gower Street, London WC1E 6BT, United Kingdom}

\date{\today}

\begin{abstract}
Interfacing between matter qubits and light is a crucial provision for numerous quantum technological applications.
However, a generic qubit may not directly interact with a relevant optical field mode, and hence, one could necessitate to adjust frequencies to match resonance conditions between parties. In this work, we show how a parametric coupling of the qubit with a mechanical oscillator, in conjunction with the trilinear radiation pressure coupling of the same object with light, can induce maximal qubit-light entanglement at an optimal time. We also show how our method enables conditional nonclassical state preparation of an optical field state via qubit measurement in the weak optomechanical coupling regime, whereas nonclassical states of the same can dynamically be achieved in the moderate-to-strong single-photon coupling limit. Our scheme benefits from not requiring any cooling of the mechanical component, and not needing an adjusting of the detunings and transition frequencies to have resonance between any pairs of quantum systems.
\end{abstract}

\maketitle

\section{Introduction}



Recent work on multipartite hybrid quantum systems attests to be a promising avenue to fulfill high-level quantum control and manipulation of quantum systems \cite{review1, review2, wallquist2009}. Essentially, such hybrid platforms serve as multitasking modular architectures; hence, each building-block (commonly with disparate frequencies) of the assembled unit will play a different role into encoding, processing, distributing, and reading out the quantum information. 

A straightforward example to observe the benefits of hybrid setups is to consider nodes allocated in a quantum network (for light-mediated qubit-oscillator indirect coupling we refer the interested reader to Refs. \cite{review2, wallquist2010}, for example). There the distribution of information between remote nodes is accomplished via photonic qubits (or propagating phonons in extended phonon waveguides \cite{Habraken2012}), whereas the encoding (or storing) can be attained within the node itself utilizing matter qubits or mechanical oscillators \cite{Kimble}. Typically, the interplay between matter-light parties demands resonant (or quasi-resonant) interactions between a qubit and a cavity field \cite{Kimble} ---this is the case, for example, for trapped single or collective two-level atoms inside a cavity \cite{restrepo2014, review2}. The qubit-cavity direct coupling could be used to either map the qubit state to a cavity field so that it was carried and fed into a distant cavity via the light \cite{Cirac97, Pellizzari, van-Enk-Kimble}, or to entangle a qubit maximally with the field in a cavity. Subsequent joint detections of the light fields from two separate cavities could be used to maximally entangle the qubits in a heralded manner \cite{Bose1999b}.

In recent years a plethora of other qubits have surfaced which have frequencies in the microwave and radio-frequency range \cite{pla2012, Maragkou2015, Kolkowitz-science, RABL, Rabl-nature}. For these, an alternate strategy has been suggested whereby both the qubit and the optics interact with a mechanical mediator, i.e., no direct coupling between the qubit and the cavity takes place. These systems have shown the successful linking of distant qubits through optomechanics \cite{stan2010, stan2011, stan2012, hab2012}. However, these schemes rely on an exchange of excitations between systems ---between the qubit and the mechanics (a Jaynes-Cummings interaction) and between the mechanics and light (a beam-splitter interaction). In the case of Jaynes-Cummings combined with beam-splitter interaction, it is intuitive that the former and latter Hamiltonians swap quantum states from the qubit to mechanics and mechanics to light respectively \cite{review1}. Schemes relying on the exchange of excitations are only ensured at the cost of an appropriate adjustment of detunings of the fields. Hence, driving the qubit and the optical field from their respective transitions by precisely the mechanical frequency. 


\begin{figure}
 \centering \includegraphics[width = 0.8 \linewidth]{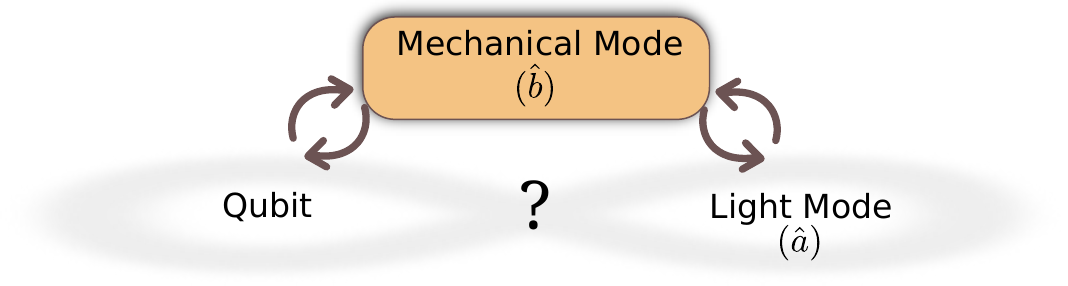}
 \caption{A single qubit interacts with a mechanical oscillator [with boson mode ($\hat{b}$)], whereas the latter object is coupled to a quantized light field ($\hat{a}$). As no direct qubit-light coupling is present but mediated through the mechanical degree of freedom. We wonder whether one could reach, dynamically and under Hamiltonian parametric interactions, a high degree of qubit-light quantum entanglement.}
 \label{modela}
\end{figure}

We wonder in this work, whether we can relax the above resonance conditions between parties, and thus, to combine distinct elements into an off-resonant hybrid system. To address this question, we make use of parametric Hamiltonians ---which are not of a state swapping type by nature. It is, therefore, relevant to explore whether parametric interactions of qubits with mechanics and the parametric trilinear optomechanical interaction can be fruitfully used to entangle a qubit with light. Specifically, we study a system, as shown in Fig. \ref{modela}, where a generic qubit becomes entangled to a cavity mode mediated through a mechanical object. Here we consider the qubit directly coupled to the mechanical oscillator position. Such a qubit-oscillator Hamiltonian can be synthesized in a variety of ways, notably through magnetic field gradients \cite{scala2013} or through capacitive couplings \cite{ARMOUR}. On the other hand, for the oscillator-cavity subsystem, we exploit the nonlinear radiation pressure interaction. Very recently, related hybrid qubit-optomechanics systems are of interest fundamentally in giving rise to interesting polaritonic states involving mechanics \cite{restrepo2014} and tripartite entanglement \cite{abdi}. Through these Hamiltonians, the most sensible task perhaps is to look for entanglement between the qubit and the optical fields. Even in this task, whether the entanglement will be ``transitive'' in nature is a priori not clear. We show here that, indirect oscillator-mediated qubit-cavity entanglement can indeed be achieved under parametric Hamiltonians without any necessity for adjusting resonances.


The rest of the article reads as follows: In Sec. \ref{tripartite-dynamics}, we focus our attention on the entanglement dynamics for several initial states both for the light field and for the mechanical oscillator. For the initial pure states considered by us, we give a closed form for the entanglement dynamics \cite{Alexandra}, which allow us to tune the interaction couplings to reach maximal qubit-cavity entanglement at a specific time. We show how to exploit the dispersive nature of our parametric Hamiltonian, where both the total number of photons, as well as the qubit excitations are conserved throughout the closed quantum dynamics. In Sec. \ref{master-eq.sec}, we devote the quantum entanglement analysis in the presence of decoherence. In general, we consider the strong coupling regime, i.e., where the relevant frequencies of our system exceed the damping rates of the open dynamics. For this study, we numerically solved the master equation \cite{qutip} in a dressed picture, as the single-photon radiation strength operates in the moderate-to-strong optomechanical regime. In Sec. \ref{sec:prepa}, we investigate the conditional preparation of nonclassical states of the optical field by i) its dynamics alone, and ii) via measurements on the qubit subsystem. Finally, in Sec. \ref{sec-conc}, we conclude and outline the main results of our work.

\section{Qubit-Optomechanical dynamics}\label{tripartite-dynamics}

Let us study a hybrid tripartite qubit-optomechanical system as sketched in Fig. \ref{modela}. In particular, we consider a generic qubit coupled dispersively (and non-resonantly) to a mechanical object, whereas a single cavity mode interacts to the latter one via trilinear radiation pressure interaction. For simplicity, we will model the quantum evolution in the absence of decoherence, in a later section we will numerically solve the open dynamics in the presence of both light and mechanical energy losses.

As the mechanical oscillator mediates the qubit and the light, we can write the Hamiltonian in a frame rotating at the spin frequency as following ($\hbar = 1$)

\begin{equation}
\hat{H}_{\mathrm{int}} = \omega_m\hat{b}^\dag\hat{b} - (g_0\hat{a}^\dag\hat{a} + \lambda_0 \hat{\sigma}_z)(\hat{b} + \hat{b}^\dag), \label{hamil}
\end{equation}

where $\omega_m$ stands for the mechanical frequency, $g_0$ is the cavity-oscillator radiation pressure coupling, and $\lambda_0$ is the qubit-oscillator coupling strength; $\hat{a}$ ($\hat{b}$) is the boson operator for the cavity (oscillator), respectively. And, $\hat{\sigma}_z$ is the Pauli $z-$matrix for the qubit. Here, we would like to stress that, from Eq. (\ref{hamil}), one may notice that the individual Hamiltonians do indeed entangle a qubit-mechanical oscillator pair and an optical field-mechanical oscillator pair, but whether also a qubit-optical field entanglement will result from this is hard to guess. 

To solve the quantum dynamics, we proceed to derive the evolution operator for the Hamiltonian in Eq. (\ref{hamil}) as done previously in Refs. \cite{sougato, montenegro, mancini}

\begin{equation}
\hat{U}(t) = e^{i(g\hat{a}^\dag\hat{a} + \lambda\hat{\sigma}_{z})^2(t - \sin t)} e^{(g\hat{a}^\dag\hat{a} + \lambda\hat{\sigma}_{z})(\eta \hat{b}^\dag - \eta^*\hat{b})}e^{-it\hat{b}^\dag\hat{b} }\label{U}
\end{equation}

where $\eta \equiv \eta(t) = 1 - e^{-it}$, and $\{g = g_0/\omega_m, \lambda = \lambda_0/\omega_m\}$ are the scaled coupling parameters.

Motivated for the indirect generation of qubit-cavity quantum entanglement, a simple inspection in Eq. (\ref{U}), suggests some directions regarding the initial states for the qubit and the cavity field. Those indications are due to the dispersive interaction, \textit{i.e}., a Hamiltonian conserving both the qubit and optical excitations. Hence, to attempt qubit-light entanglement generation, one requires to initialize the qubit (cavity) different from an eigenstate of $\hat{\sigma}_z$ (Fock number state). Otherwise, no entanglement between the qubit (or the cavity) can be generated with the rest of the subsystems, as they persist disentangle during the evolution. For the sake of simplicity, we will consider throughout this work an initial qubit superposition $1/\sqrt{2}(\ket{\uparrow} + \ket{\downarrow})$, where several initial states for the cavity state and the mechanical object will be studied.

\subsection{Optical qubit and mechanical oscillator in coherent state}

Let us evolve the system from an initial superposition of Fock number states $1/\sqrt{2}(\ket{0} - \ket{1})$ for the cavity field \cite{sougato01} (optical qubit). Although easy to formulate mathematically, this state remains difficult to prepare experimentally. Preparation techniques of the optical mode in a state $\sim \ket{0} - \ket{n}$ have been reported previously \cite{Marte, Vogel, Kimblen}, being $n=1$ the most simple state to generate in the laboratory. On the other hand, we initialize the mechanical object in a coherent state with an amplitude of $\beta > 0$. With these initial conditions, we intend to unravel the tripartite dynamics from states with low Hilbert spaces; more available initial states are considered in later sections. By evoking the unitary operator from Eq. (\ref{U}), it is straightforward to obtain the following wave function:

\begin{widetext}
\begin{eqnarray}
 \nonumber \ket{\psi(t)} &=& \frac{1}{2}\Big[ e^{i\lambda^2(t - \sin t)} e^{i\lambda\beta\sin t}\up\ket{0}\ket{\beta e^{-it} + \lambda\eta} - e^{i(g+\lambda)^2(t - \sin t)} e^{i(g+\lambda)\beta\sin t}\up\ket{1}\ket{\beta e^{-it} + (g+\lambda)\eta}\\
 &+& e^{i\lambda^2(t - \sin t)} e^{-i\lambda\beta\sin t}\down\ket{0}\ket{\beta e^{-it} - \lambda\eta} - e^{i(g-\lambda)^2(t - \sin t)} e^{i(g-\lambda)\beta\sin t}\down\ket{1}\ket{\beta e^{-it} + (g-\lambda)\eta}\Big].\label{evolution_fock}
\end{eqnarray}
\end{widetext}

To investigate the quantum correlations of the qubit-optomechanical wave function in Eq. (\ref{evolution_fock}), we proceed to calculate the quantum entanglement between bipartite systems. In general, throughout this work the entanglement will be mainly computed using the negativity [$\mathcal{N}(t)$] \cite{neg2}, a quantity defined as

\begin{equation}
 2 \mathcal{N}(t) = \sum |\varepsilon_i| - \varepsilon_i
\end{equation}

where $\varepsilon_i$ are the eigenvalues of the partially transposed reduced density matrix at fixed time $t$. 

\begin{figure}
  \centering \includegraphics[width = \linewidth]{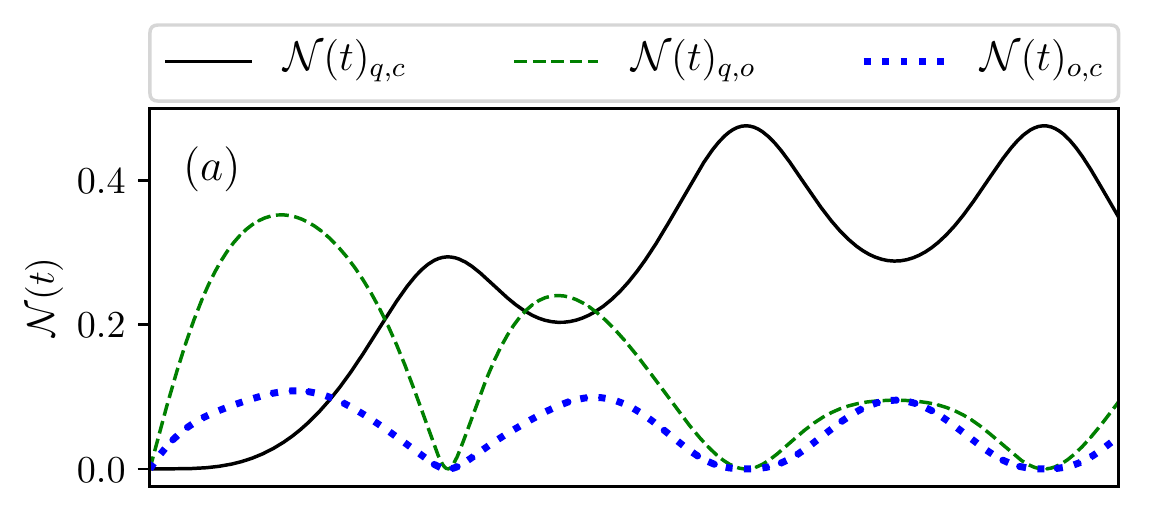}
  \centering \includegraphics[width = \linewidth]{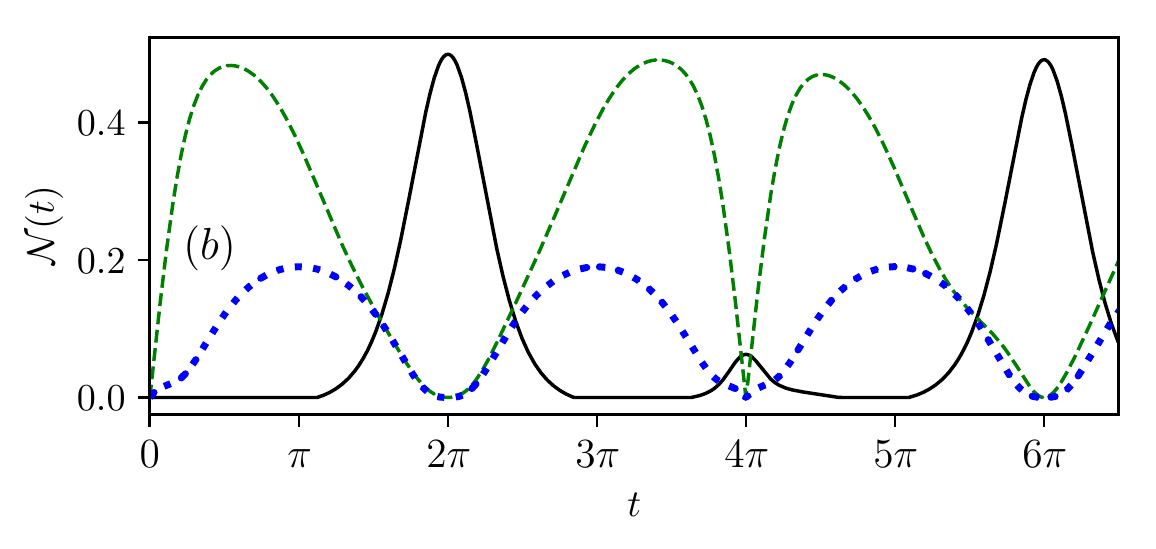}
 \caption{(Color online) Nagativity dynamics [$\mathcal{N}(t)$] for different reduced density matrices (bipartite states). We plot the qubit-cavity [$\mathcal{N}(t)_{q,c}$, solid line], the qubit-oscillator [$\mathcal{N}(t)_{q,o}$, dashed line], and the oscillator-cavity negativity [$\mathcal{N}(t)_{o,c}$, dotted line] for different qubit-optomechanical coupling values. Top (Bottom) panel considers  $g = 0.2$ and $\lambda = 0.25$ ($g=0.2$ and $\lambda = 0.625$); $\beta = 1$.}\label{fig:negativity_fock}
\end{figure}

In Fig. (\ref{fig:negativity_fock}) we illustrate the dynamics of quantum entanglement for a set of qubit-optomechanical coupling values $\{g, \lambda\}$. Concretely, we have computed the qubit-cavity [$\mathcal{N}(t)_{q,c}$], the qubit-oscillator [$\mathcal{N}(t)_{q,o}$], and the oscillator-cavity negativity [$\mathcal{N}(t)_{c,o}$]. In the top panel (a) of Fig. (\ref{fig:negativity_fock}), we chose $g = 0.1$ and $\lambda = 0.4$ (for $\beta = 1$). As seen, the mechanical oscillator disentangles from the rest of the subsystems at each cycle, i.e., at times $2\pi/\omega_m\times l$ ($l$ being an integer), the qubit-oscillator (and cavity-oscillator) negativity vanishes to zero ---this can be seen as $\eta(t=2\pi) = 0$, and therefore, $\ket{\beta e^{-2\pi i} \pm \lambda \eta} = \ket{\beta e^{-2\pi i} + (g \pm \lambda) \eta} = \ket{\beta}$. On the contrary, the qubit-cavity negativity at those very same times has a non-zero value, reaching its first maximum of $\mathcal{N}(t)_{q,c} = 0.5$ at $t = 4\pi$. Thus, the initially disentangled qubit-cavity subsystems have been indirectly (and maximally) entangled through a mechanical object.

Furthermore, one may require to possess highly entangled states at the earliest in the quantum dynamics, for instance, to avoid detrimental effects due to decoherence or for quantum computing/processing purposes. To reach the maximum qubit-cavity negativity at $t = 2\pi$, we need to optimize the set of qubit-optomechanical parameters $\{g, \lambda\}$ in the negativity function. However, albeit quite manageable to compute numerically, a closed analytical form is usually difficult to obtain. To overcome this obstacle, we proceed to derivate a simple form encompassing several partitions of the tripartite system. This simple, yet rich procedure \cite{Alexandra}, shows that an appropriate addition (subtraction) of individual entropies should be capable of quantifying the degree of entanglement within each of the sub-systems. The intrinsic qubit-cavity entanglement is defined as follows:

\begin{equation}
\mathcal{E}_{q,c}(t) = \mathcal{S}(t)_q + \mathcal{S}(t)_c - \mathcal{S}(t)_o.
\end{equation}

In the above, $\mathcal{S}(t)_i = 1 - \mathrm{Tr}[\hat{\rho}^2_i(t)]$ is the linear entropy, and $\hat{\rho}_i(t) = \mathrm{Tr}_{j,k}\hat{\rho}_{i,j,k}$ is the corresponding reduced density matrix. With the above definition, the intrinsic qubit-cavity entanglement is reduced to:

\begin{flalign}
 \nonumber &\mathcal{E}_{q,c}(t) = \frac{1}{8} \Big[e^{2 (g+2 \lambda)^2 (\cos t-1)}+e^{2 (g-2 \lambda)^2 (\cos t-1)}+2&&\\
 &- 2 \left[e^{2 g^2 (\cos t-1)}+e^{8 \lambda^2 (\cos t-1)}\right] \cos [4 g \lambda (t-\sin t)]\Big],&&
\end{flalign}
 
where, for the special case of $t = 2\pi$ simplifies to

\begin{equation}
  \mathcal{E}_{q,c}(t=2\pi) = \sin^2(4g\lambda\pi). \label{intrinsic-1}
\end{equation}

From the above equation, we can readily notice that $\mathcal{E}_{q,c}(t=2\pi)$ reaches its first maximum at any couplings combinations given by $g\lambda = 1/8$. In Fig. \ref{fig:negativity_fock}-(b) we model this case, where we have considered $g = 1/5 = 0.2$ and $\lambda = 5/8 = 0.625$. Any tuple $g\lambda = 1/8$ would deliver maximal entanglement. Nevertheless, as $g$ decreases to the weak-to-moderate radiation pressure regime, one needs $\lambda$ to increase to the strong spin-mechanical regime. On this matter, we would like to notice that both values are within the current experimental feasibility. Vast efforts to increase the single-photon radiation pressure coupling have been presented during the last years. For example, the above moderate optomechanical operational regime of $g = g_0/\omega_m \lesssim 0.2$ have been reported in quantum cavity pulsed optomechanics \cite{vanner} and other setups \cite{Painter,chan,cooling1,murch,xuereb}. 


As seen from this section, we have fully accomplished the indirect qubit-cavity entanglement mediated through a mechanical object. However, even though an initial optical qubit gives us insight into the qubit-optomechanical evolution and the entanglement dynamics, this particular preparation for the light field is hard to access experimentally. For that reason, in the next section we will study the merits and demerits of having a more feasible preparation for the optics; a coherent state.

\subsection{Initial coherent states for both the light and the mechanics}

Assume that initially both the mechanics and the optics fields are in a coherent state preparation, $\ket{\beta}$ and $\ket{\alpha}$, respectively. On the one hand, to generate such a state for the light, one can notice that once the cavity has no intracavity photons, it could be pumped by an external laser, and therefore, driving the vacuum state towards a displaced vacuum state $\ket{\alpha}$. This driving process is realized on a time-scale which is much shorter than the time-scale of the oscillator's motion, and therefore not having a significant perturbation in the mechanical dynamics. And, similarly for a mechanical  coherent state. Several theoretical and experimental techniques have been proposed to reach the ground state cooling for mechanical oscillators, and even recently demonstrated cooling of mechanical vibrational modes close to the quantum ground state \cite{oconell}.

The initial quantum state $\ket{\psi(0)} = 1/\sqrt{2}(\up + \down)\ket{\alpha}\ket{\beta}$ then evolves according to:

\begin{eqnarray}
 \nonumber \ket{\psi(t)} = \sum_{n=0}^\infty(C_n^+ (t)\ket{\uparrow}\ket{\phi_n^+(t)} + C_n^-(t)\ket{\downarrow}\ket{\phi_n^-(t)})\ket{n}\label{psit}\\
\end{eqnarray}

where,

\begin{equation}
 C_n^\pm(t) = \frac{\alpha^n}{\sqrt{2n!}}e^{-|\alpha|^2/2}e^{i(gn \pm \lambda)^2(t - \sin t)}e^{i(gn \pm \lambda)\mathrm{Im}(\eta \beta)}
\end{equation}

and $\phi_n^\pm(t) = \beta e^{-it} + (gn \pm \lambda)\eta$. 

In contrast to our previous section, where the optics were restricted to a two-level system --- being zero and one photons. Here, the explicit optomechanical Kerr-like term $e^{i(gn \pm \lambda)^2(t - \sin t)}$ in $C_n^{\pm}(t)$ plays a significant role in the dynamics. On the other hand, each qubit eigenstate dynamically couples to different coherent amplitudes $\ket{\phi_n^\pm(t)}$, as each qubit component effectively shifts the mechanical potential proportional to $\lambda_0$ ---a shift also found in the previous section.

To explore further the bipartite dynamics, we proceed to derive the reduced density matrices as following:

\begin{flalign}
&\hat{\rho}_{o,c}(t) = \sum_{\substack{n,m=0 \\ j=\{+,-\}}}^\infty C_n^j(t)C_m^{j*}(t)\ket{n, \phi_n^j}\bra{m,\phi_m^j},\label{mdr-1}&&\\
 \nonumber &\hat{\rho}_{q,o}(t) = \sum_{n=0}^\infty C_n^+(t)C_n^{+*}(t)\ket{\uparrow, \phi_n^+}\bra{\uparrow, \phi_n^+} + C_n^+(t)&&\\
 \nonumber &\times C_n^{-*}(t)\ket{\uparrow, \phi_n^+}\bra{\downarrow, \phi_n^-} + C_n^-(t)C_n^{+*}(t)\ket{\downarrow, \phi_n^-}\bra{\uparrow, \phi_n^+}&&\\
 &+ C_n^-(t)C_n^{-*}(t)\ket{\downarrow, \phi_n^-}\bra{\downarrow, \phi_n^-},\label{mdr-2}
\end{flalign}

being the oscillator-cavity (o,c) and the qubit-oscillator (q,o) bipartite subsystems, respectively. And, 

\begin{flalign}
\nonumber &\hat{\rho}_{q,c}(t) = \sum_{n,m=0}^\infty \Big(C_n^+(t)C_m^{+*}(t)\phi_{mn}^{++}\ket{\uparrow}\bra{\uparrow} + C_n^+(t) &&\\
\nonumber &\times C_m^{-*}(t)\phi_{mn}^{-+}\ket{\uparrow}\bra{\downarrow} + C_n^-(t)C_m^{+*}(t)\phi_{mn}^{+-}\ket{\downarrow}\bra{\uparrow}&&\\
&+ C_n^-(t)C_m^{-*}(t)\phi_{mn}^{--}\ket{\downarrow}\bra{\downarrow}\Big)\otimes\ket{n}\bra{m}\label{mdr-3}
 \end{flalign}

stands for the qubit-cavity (q,c) reduced system; we defined $\phi_{mn}^{ij} = \bra{\phi_m^i}\phi_n^j\rangle$.

In analogy with the previous section, it is straightforward to notice that for each cycle of the mechanical object ($t = 2\pi \rightarrow \eta = 0$), the oscillator disentangles from the cavity as well as the qubit state. It is worthy of expressing the wave function at such particular time

\begin{eqnarray}
 \nonumber \ket{\psi(t = 2\pi)}_{q,c} &=& e^{-|\alpha|^2/2}\sum_{n=0}^\infty \frac{\alpha^n}{\sqrt{2n!}}\Big[e^{i(gn + \lambda)^22\pi}\up \\
 &+& e^{i(gn - \lambda)^22\pi}\down\Big]\ket{n}. \label{coinciding}
\end{eqnarray}

\begin{figure}
 \centering \includegraphics[width = \linewidth]{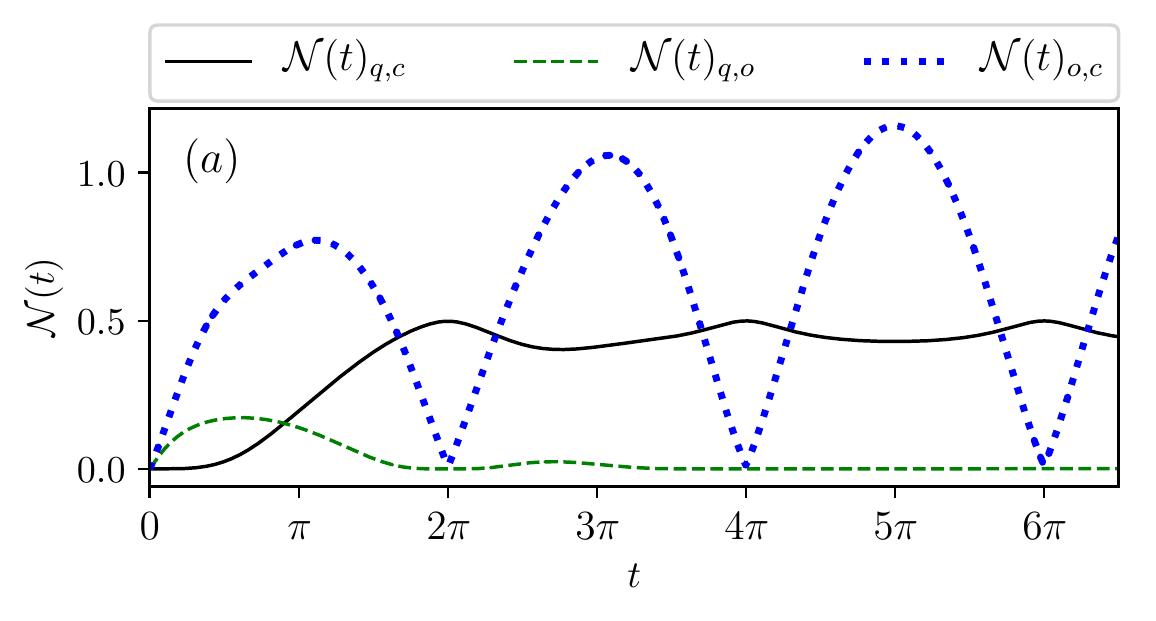}
 \centering \includegraphics[width = \linewidth]{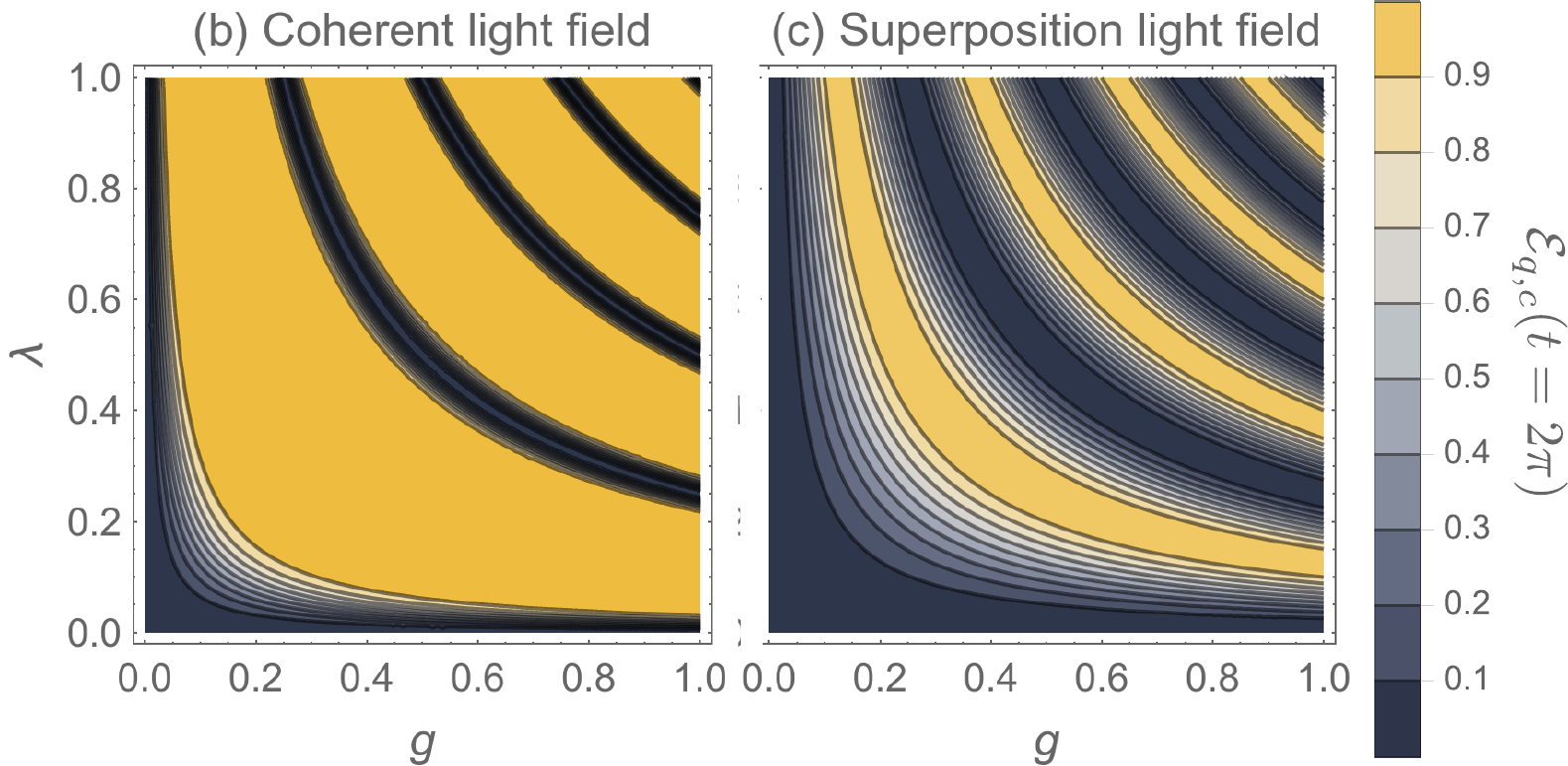} 
 \caption{(Color online) (a) Different bipartite entanglement dynamics for an initial cavity and mechanical coherent state. In (b), we plot the intrinsic entanglement for this case [Eq. (\ref{intrinsic-eq})]. And, in (c) we show the same for Eq. (\ref{intrinsic-1}). Other values are $\beta = 2, \alpha = 2, g = 0.2, \lambda = 0.25$}\label{fig:negativity_coherent}
\end{figure}

We investigate the entanglement dynamics for each bipartite system in Fig. \ref{fig:negativity_coherent}. Two clear advantages when compared to the previous case arises, i) the qubit-cavity entanglement reaches its maximum faster than in the optical qubit scenario, and ii) the qubit-cavity entanglement does not decrease nor drop to zero for a larger time window. Other values are $\beta = 2, \alpha = 2, g = 0.2, \lambda = 0.25$

To find a suitable set of qubit-optomechanical couplings such as maximizes the qubit-cavity entanglement, we follow the same procedure as in the previous section. The linear entropies can be obtained directly from the reduced density matrices already expressed in Eqs. (\ref{mdr-1}) to (\ref{mdr-3}). The intrinsic qubit-cavity entanglement at $t = 2\pi$ for the coherent case reads as:

\begin{eqnarray}
 \nonumber \mathcal{E}_{q,c}(t=2\pi) &=& 1 - \sum_{\substack{n=0\\m=0}}^\infty \frac{e^{-2 \alpha^2} \alpha^{2 (m+n)} \cos [8 \pi  g \lambda (m-n)]}{m! n!}\\
 &=& 1 - e^{-4\alpha^2\sin^2[4g\lambda\pi]}. \label{intrinsic-eq}
\end{eqnarray}

Notice from the above Eq. (\ref{intrinsic-eq}) that $\sin(4g\lambda\pi)$ appears as in the previous section. However, while for an initial optical qubit the condition $4g\lambda\pi = \pi/2$  gives rise directly to a maximal entanglement, for the coherent case the same condition drives asymptotically the intrinsic entanglement towards unity. We compare both situations in the bottom panels of Fig. \ref{fig:negativity_coherent}.

One could also examine the entanglement dynamics for different operational regimes. Firstly, let us consider the case when the single-photon radiation pressure coupling dominates over the spin-mechanical coupling, i.e., scaled frequencies $g \gg \lambda$. Basically, in this regime, where the optomechanical coupling dominates, one can observe (for a large number of mechanical cycles $t \gg 1$) that the tripartite system evolves mainly as a coherent evolution between the light and the mechanics. And, interestingly, between the qubit and the cavity field ---disregarding (in a sense) the spin-mechanical interaction. In the opposite case when the spin-mechanics interaction dominates, i.e., $g \ll \lambda$. We find that the qubit entangles the mechanical object, whereas, at larger times, a coherent dynamics between the qubit-cavity system is also achieved. To successfully attain a high indirect qubit-cavity entanglement the optimal operational regime is when both $\{g, \lambda\}$ are comparable with the mechanical frequency $\omega_m$. Under these values, we can reach higher qubit-cavity quantum entanglement at faster mechanical cycles in the dynamics. Notice that, it is also possible to achieve ample time windows where the qubit-cavity entanglement can oscillate between its maximum $\sim 0.5$ for values such as $g \ll 1$ and $\lambda \sim 1$, however, this is only possible far from the transient dynamics, i.e., $t \gg 1$ ---a time domain not being under consideration in this work.

\subsection{Coherent state for the cavity field and a thermalized oscillator}

A step forward to study the bipartite entanglement under different initial preparations is to consider an initial coherent state for the light field and a mechanical oscillator at temperature $T$. In principle, as we intend to utilize the mechanical object solely as a mediator, i.e., not having a direct access to the mechanics throughout the quantum evolution nor any measurement is being performed on this object, it is more precise to consider an initial thermal state representation for the mechanical subsystem. Thus, let us consider in this section that the cavity state is being prepared into a coherent state as before, whereas the mechanical object will be considered as a thermal state at temperature $T$, i.e., $\hat{\rho}(0)^{th} = 1/\pi\bar{n}\int \ket{\beta}\bra{\beta}\mathrm{exp}(-|\beta|^2/\bar{n})d^2\beta$, where $\bar{n} = [\mathrm{exp}(\hbar\omega_m/k_BT) - 1]^{-1}$ stands for the thermal occupancy number, and $k_B$ is the Boltzmann constant. By solving the Schr\"{o}dinger equation we can readily get the tripartite normalized density matrix as following:

\begin{eqnarray}
 \nonumber \hat{\rho}(t) &=& \sum_{n,m=0}^{\infty} \ket{n} \bra{m} \otimes \Big[\up  \updag \otimes \hat{\rho}^{th}_{++} + \down\downdag \otimes \hat{\rho}^{th}_{--}  \\
 &+& \up\downdag \otimes \hat{\rho}^{th}_{+-} + \down \updag \otimes \hat{\rho}^{th}_{-+}\Big].\label{evolution}
\end{eqnarray}

where, we have simplified the notation as $\hat{\rho}^{th}_{ab} \equiv \hat{\rho}^{th}_{ab}(n,m,t) = 1/(\pi\bar{n}) \times \int C^a_n(t) C^{*b}_m(t)\ket{\phi_a^n(t)}\bra{\phi_b^m(t)} e^{-|\beta|^2/\bar{n}}d^2\beta$, where $\{a,b\}$ might be $+$ or $-$.

From Eq. (\ref{evolution}),  we can notice that the mechanical object becomes disentangled at each mechanical cycle once again. Nonetheless, when compared to previous cases where the tripartite state remains pure throughout the dynamics, here, it is not a direct consequence. Moreover, the qubit-cavity reduced bipartite undergoes from the initial mixed state to a pure state at each oscillator's cycle. And, it coincides with the above wave function derived in the previous section [see Eq. (\ref{coinciding})], being independent also from the thermal occupancy number $\bar{n}$

\begin{eqnarray}
 \nonumber \ket{\psi(t = 2\pi)}_{q, c} &=& e^{-|\alpha|^2/2}\sum_{n=0}^\infty \frac{\alpha^n}{\sqrt{2n!}}\Big[e^{i(gn + \lambda)^22\pi}\up \\
 &+& e^{i(gn - \lambda)^22\pi}\down\Big]\ket{n}. \label{coin1}
\end{eqnarray}

Notwithstanding the coinciding wave functions in Eq. (\ref{coinciding}) and Eq. (\ref{coin1}), the scenario involving an evolved thermal mechanical oscillator diverges from the coherent state in the transient intervals. For instance, let consider the time interval $2\pi < t < 10\pi$ and $\bar{n} = |\beta|^2 = 4$ (to relate as in the previous section), at this particular time window the oscillator becomes disentangle in the same amount as for the coherent case for times multiples of $t = 2\pi$, however, quantum entanglement oscillating near its maximum value [as in Fig. \ref{fig:negativity_coherent}-(a)] was not reported for any set of qubit-optomechanical couplings strengths in the domain $0 < \{g, \lambda\} \leq 1$. In general, the mixedness of the oscillator conduces to diminish the entanglement between the qubit-cavity parties, only when the mechanical object is highly disentangled from the rest of the system a maximal qubit-cavity entanglement conditioned on $g$ and $\lambda$ can be generated. To study the impact of $\bar{n}$ on the entanglement dynamics, we simulate (numerically) the evolution for the range $0 \leq \bar{n} \leq 20$. In the domain of low phonons on average, let say $\bar{n} \leq 5$, the qubit-cavity entanglement shows to be quasi-stabilized for some sets of qubit-optomechanical values (in agreement with the reduction of mechanical mixedness). However, this behavior emerges for larger times when compared to the coherent dynamics, where entanglement between $2\pi$-peaks tending to decrease as $\bar{n}$ increases.

\section{dynamics in the presence of losses}\label{master-eq.sec}

Quantum correlations, or ultimately quantum coherence, generally suffer from a decrement when the relevant system of interest evolves in contact with their surroundings (i.e., an open quantum evolution). In the absence of an engineered reservoir or suitable dissipative mechanism, quantum correlations are a frail resource in the presence of noise. In this section, we will study the effects of the impact of energy losses on the quantum dynamics, mainly, centering our attention on the qubit-cavity subsystems. The resulting detrimental effects on the unitary evolution stems from several decoherence channels, wherein for our particular qubit-optomechanical scheme we will consider energy losses arising from each element. As shown in the previous section, to achieve maximal qubit-cavity entanglement at $2\pi$ mechanical oscillations, moderate-to-strong optomechanical radiation pressure interaction must be attained. Within the single-photon strong or ultrastrong optomechanical coupling regime ($g$ and $\lambda$ is comparable to the mechanical frequency $g \sim 1$), photons and electronic states have been found to been dressed by phonon excitations of the mechanical mode strongly. And, therefore, when $g$ and $\lambda$ operates in this limit one necessarily needs to consider that the single-photon optomechanical coupling modifies the eigenstates of the system \cite{dressed-master-equation}. The corresponding optomechanical master equation in the optomechanical dressed picture (DME), together with the dressed Lindbladian for the qubit element is:

\begin{flalign}
 \nonumber &\frac{d}{dt}\hat{\rho}(t) = -i[\hat{H}, \hat{\rho}(t)] + \gamma_m(n_\mathrm{th} + 1)\mathcal{L}[\hat{b} - g \hat{a}^\dagger \hat{a}]\hat{\rho}(t) + &&\\
 \nonumber &\kappa\mathcal{L}[\hat{a}]\hat{\rho}(t) + \gamma_m n_\mathrm{th} \mathcal{L}[\hat{b}^\dagger - g \hat{a}^\dagger \hat{a}]\hat{\rho}(t) + \Gamma(1 + n_q)\mathcal{L}[\hat{\sigma}^-] &&\\
 &+ \Gamma n_q \mathcal{L}[\hat{\sigma}^+] + \frac{\gamma_\phi}{2}\mathcal{L}[\hat{\sigma}_z] + \frac{4 \gamma_mg^2}{\ln( \frac{1 + n_\mathrm{th}}{n_\mathrm{th}})} \mathcal{L}[\hat{a}^\dagger \hat{a}]\hat{\rho}(t),\label{dressed-me}
\end{flalign}

where $\hat{H}$ corresponds to the Hamiltonian in Eq. (\ref{hamil}), and the Lindblad superoperator term

\begin{equation}
\mathcal{L}[\hat{O}]\hat{\rho} = 2\hat{O}\hat{\rho}\hat{O}^{\dagger} - \hat{\rho}\hat{O}^{\dagger}\hat{O}-\hat{O}^{\dagger}\hat{O}\hat{\rho},
\end{equation}

takes into account the dissipative mechanisms to a thermal reservoir with occupation number $n_\mathrm{th}$ to the mechanical bath and scaled (by the mechanical frequency) photon (phonon) decay $\kappa$ $(\gamma_m)$. Notice that, as we are operating in optical frequencies we have neglected its corresponding occupation number $n_{c} \ll 1$, i.e., only downwards transitions will take place $\kappa \mathcal{L}[\hat{a}]\hat{\rho}(t)$ in the optical energy ladder. Qubit decoherence channels are described by the qubit relaxation $\Gamma$ and the dressed qubit dephasing $\gamma_\phi\ = \Gamma_\phi\ + 4 \gamma_m\lambda^2 / \ln[\frac{1 + n_\mathrm{th}}{n_\mathrm{th}}] $, where $\Gamma_\phi $ is the qubit pure dephasing rate. We have also considered a common thermal reservoir for the composite system, i.e., each element of our qubit-optomechanical setup are in contact with the same environment at the same temperature, thus as $\lambda \sim 1$, we consider that $n_q = n_\mathrm{th}$.

The DME (which is also derived within the Born-Markov approximation) found in Ref. \cite{dressed-master-equation} stands as a more general case when compared to the standard master equation (SME). One can transition between DME towards SME by simply considering $g \ll 1$, consequently the joint optomechanical decoherence channel $\mathcal{L}[\hat{b} - g\hat{a}^\dagger \hat{a}]$ can be effectively approximated to $\mathcal{L}[\hat{b}]$. In what follows, we analyze the effects of noise over our relevant hybrid system in the strong coupling regime, i.e., all the relevant frequencies $\{g, \lambda\}$ are higher than the damping rates $\{\gamma, \kappa, \Gamma, \gamma_\phi\}$.

\begin{figure}[t]
 \includegraphics[width = \linewidth]{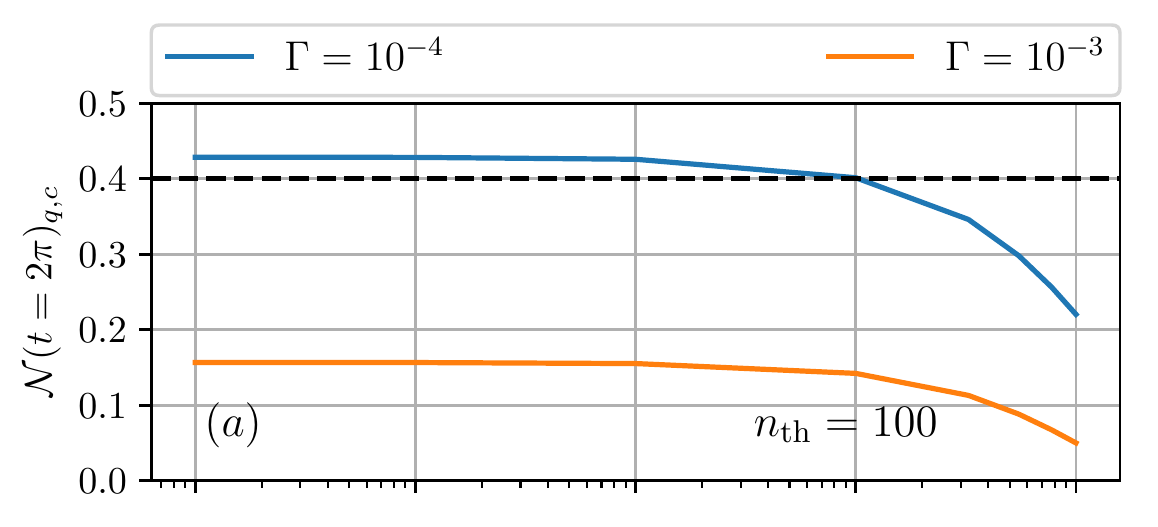}
 \includegraphics[width = \linewidth]{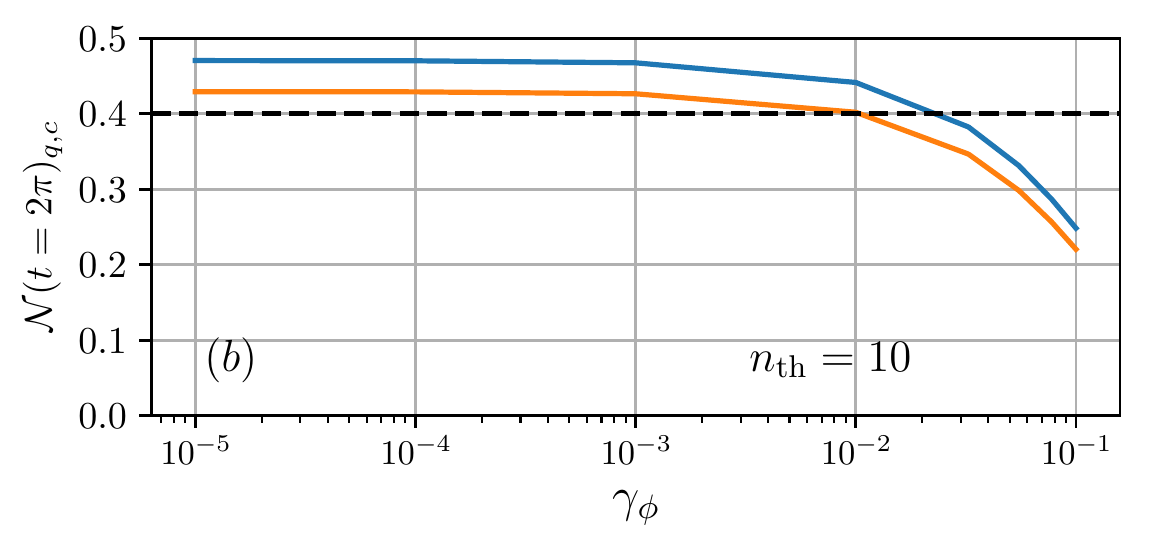}
 \caption{Qubit-cavity entanglement at $t = 2\pi$ for different spin relaxation rates $\Gamma$ as a function of the spin dephasing rate $\gamma_\phi$. We fixed $\gamma_m = 10^{-5}, \kappa = 10^{-2}$, other values as in Fig. \ref{fig:negativity_coherent}.}\label{fig:master-DME}
\end{figure}

It is typical of current optomechanical systems that the primary decoherence channel is related to the leakage of intracavity photons from the cavity, while oscillator energy losses can be second-placed in the decoherence hierarchy. For this reason, we consider a mechanical oscillator with high mechanical quality factor $Q = 10^5$ (according to our definitions this translates in $\gamma = 10^{-5}$). On the other hand, we will fix the photons decay rate equal to $\kappa = 0.01$ \cite{hybrid}. The rest of the parameters are varied in Fig. \ref{fig:master-DME}, for which we intend to illustrate up to which values of $\{\gamma_\phi, \Gamma\}$ the negativity at $t = 2\pi$ can be accommodated. The particular case considered by us is the lossless scenario depicted in Fig. \ref{fig:negativity_coherent}, where both the optics and the mechanics are initialized in coherent states. In Fig. \ref{fig:master-DME}, we arbitrarily choose a 20\% attenuation of the maximal value of the entanglement achieved in the lossless case, i.e., a detrimental tolerance up to 0.4 (negativity threshold in black dashed line). Notice that, albeit no cooling of the mechanical oscillator is truly obliged for our scheme to work, the thermal reservoir impacts on the qubit coherence. The parametric Hamiltonian, due to its qubit and photonic dispersive nature, demands to have qubit coherence to reach non-zero entanglement between these parties. In other words, no fully mixed state neither a $\hat{\sigma}_z$ eigenstate will entangle dynamically with the light field. Under the assumption of a common reservoir, $n_\mathrm{th} = 100$ [$n_\mathrm{th} = 10$] will require to have dephasing rates of $10^{-2}$, and qubit relaxing rates as low as $\Gamma = 10^{-4}$ [$\Gamma = 10^{-3}$]. 

\section{Optical nonclassical states generation}\label{sec:prepa}

In this section, we present how nonclassical states of the light field can be generated due to its dynamics alone, as well as by performing measurements on the qubit state. The tripartite platform, in this matter, may constitute an advantage in terms of light state production when contrasted to the solely optomechanical case, as no measurement over the mechanical object is being performed. The purpose of the following section intends to be primarily illustrative, and thus, we think that several other optical nonclassical states can be attained and not readily covered in the present section.

As said above, we can distinguish two cases, namely i) the generation of the nonclassical state by evolution alone, and ii) by observing the qubit subsystem. The first situation can be readily obtained from the reduced density matrix of the cavity field [obtained from Eq. (\ref{coinciding})]. At the particular time of $2l\pi$ ($l$ being an integer), the corresponding mixed state reads as:

\begin{eqnarray}
 \nonumber \hat{\rho}_c(t = 2l\pi) &=& e^{-|\alpha|^2}\sum_{n,m=0}^\infty \frac{\alpha^{n+m}}{\sqrt{n!m!}}e^{2ig^2l\pi(n^2-m^2)}\\
 &\times&\cos[4\pi gl\lambda(m-n)]\ket{n}\bra{m}.
\end{eqnarray}

\begin{figure}[t]
 \includegraphics[scale = 0.45]{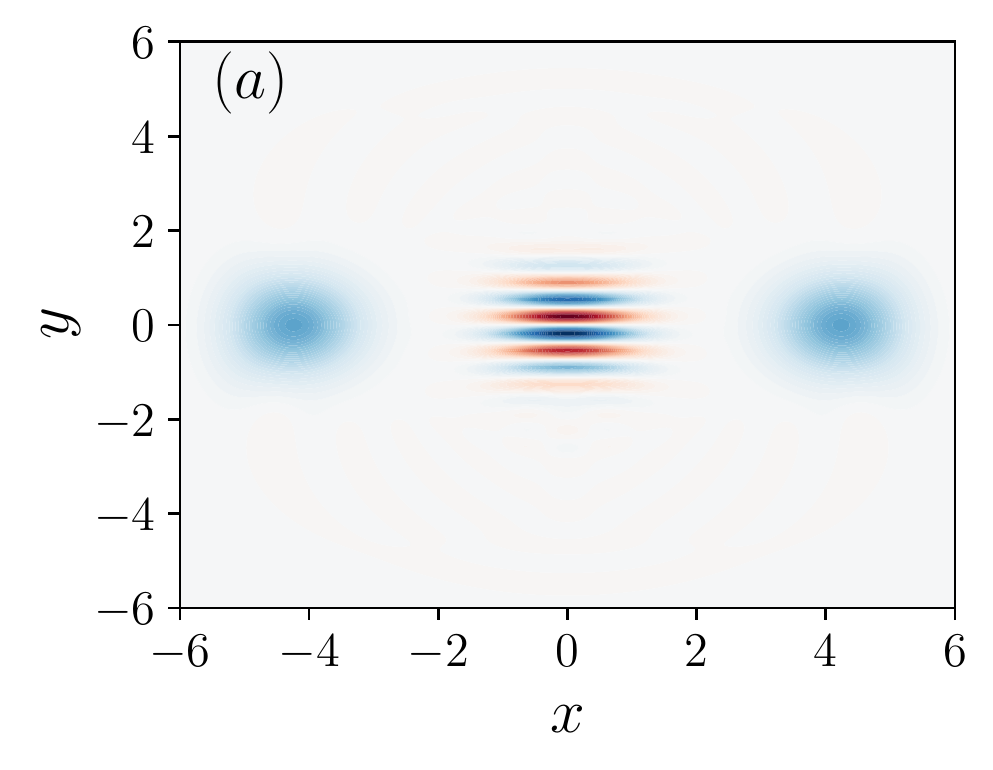}
 \includegraphics[scale = 0.45]{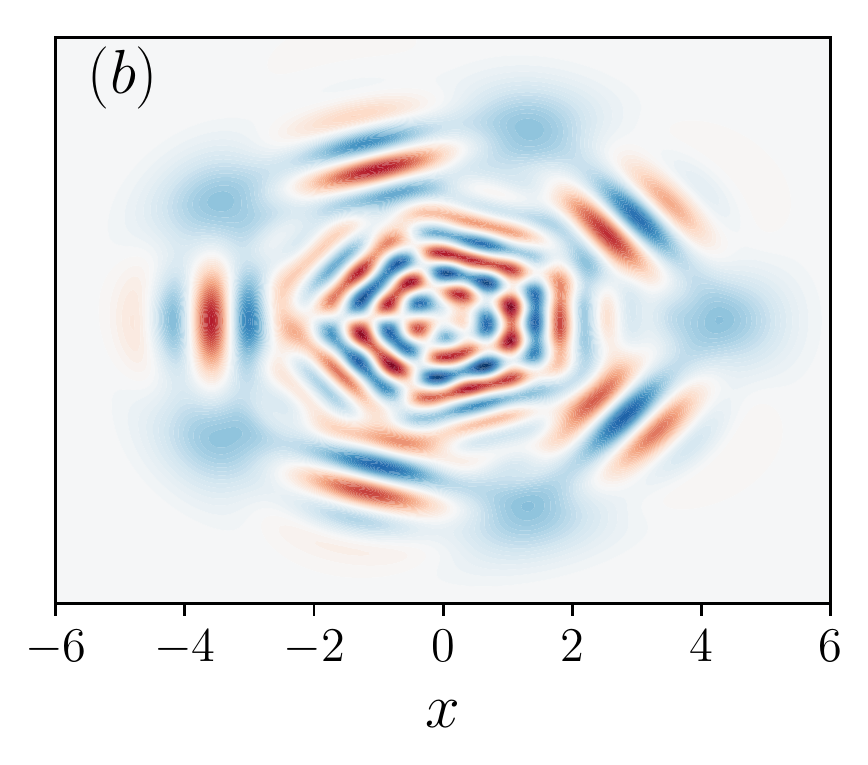} 
 \includegraphics[scale = 0.45]{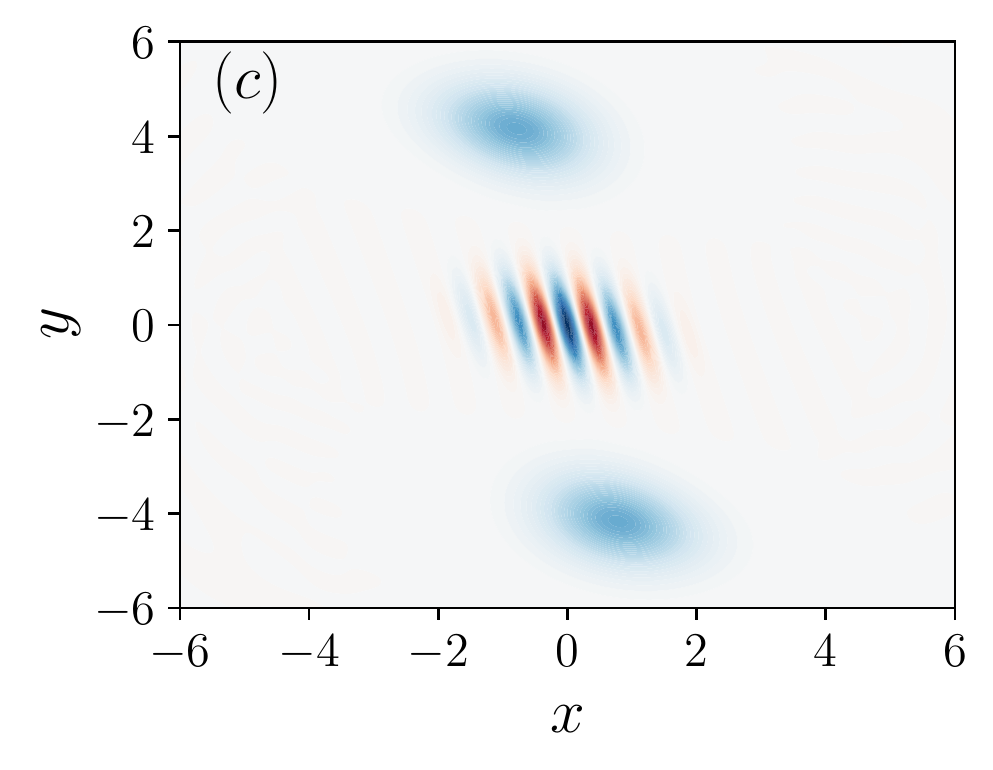}
 \includegraphics[scale = 0.45]{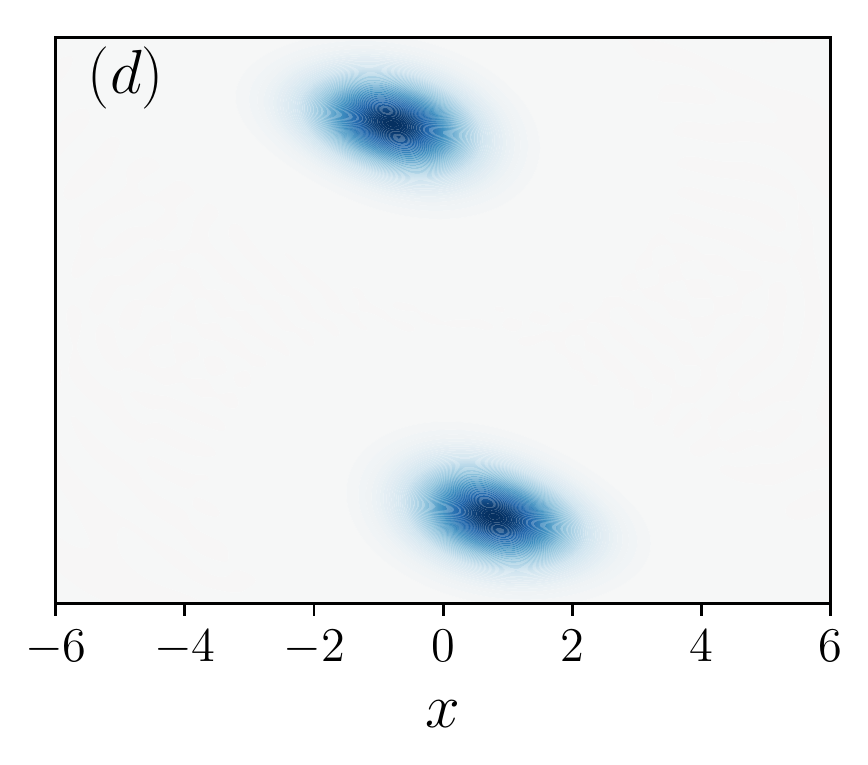} 
 
 \includegraphics[scale = 0.6]{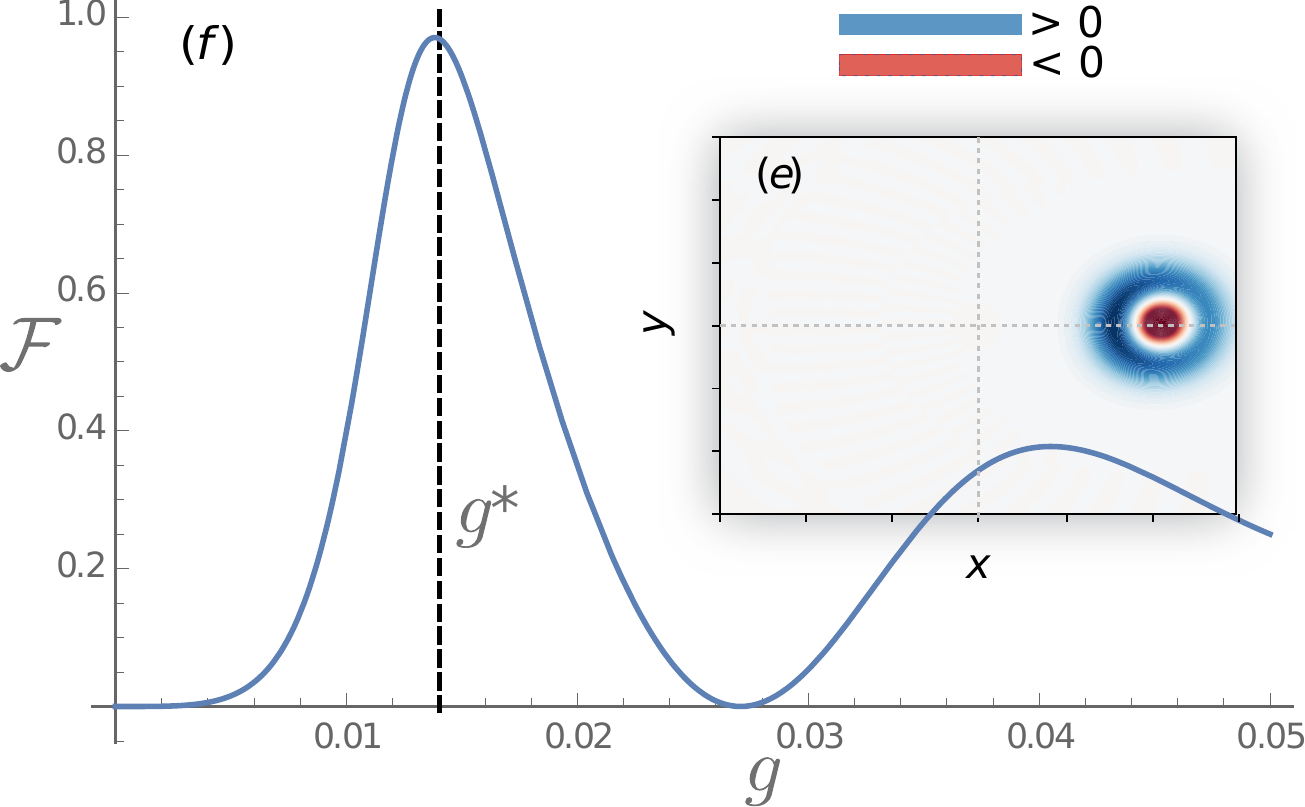}
 \caption{Panels (a) and (b) show the Wigner distribution for the generation of nonclassical states for the light field due to its dynamics alone at $t = 2\pi$. In (a) we show the two-component cat state for $g = \lambda = 0.5$, whereas (b) corresponds to the five-component cat state for $g \sim 0.32$ and $\lambda \sim 0.8$; (c) illustrates the preparation of nonclassical states via qubit projection $\ket{+}$ at $t = 10\times 2\pi$ when operating in the weak optomechanical regime $g \sim 0.013$ and qubit-mechanical coupling $\lambda = 1$. In (d) we show the impossibility to obtain a nonclassical state with the same parameters used in (c) by its dynamics alone. In panel (e) we show the displaced Fock number state $\hat{D}(\alpha)\ket{1}$ via qubit projection with its corresponding fidelity shown in (f). Other values are $\alpha = 3$, and $g^*$ the optimal optomechanical strenght to achieve $\hat{D}(\alpha)\ket{1}$. In all panels, positive (negatives) values in the Wigner function are blue (red) colored.}\label{fig:p-cats}
\end{figure}

For a particular set of values of the scaled coupling parameters $g$ and $\lambda$, we can generate nonclassical states for the cavity mode, the so-called multicomponent Schr\"{o}dinger cat states. The appropriate choice of $\{g,\lambda\}$ (when each party becomes disentangled from the rest) can be obtained from the intrinsic entanglement expression derived previously in Eq. ({\ref{intrinsic-eq}}), being $4g\lambda l = k$ ($\{k,l\}$ integers). The multicomponent Schr\"{o}dinger cat states are achieved for $g\sqrt{2 l p} = 1$, where $p \geq 2$ gives the $p-$mode Schr\"{o}dinger cat state generated. Without loss of generality, let us fix $\{l, k\} = 1$, as different values of those will involve only rotations in the phase space of the light field. In Figs. \ref{fig:p-cats}-(a) and \ref{fig:p-cats}-(b) we show the Wigner quasi-probability distributions [defined as $W(x,y) = \int_{-\infty}^\infty\left<x + x'\left| \hat{\rho}_{c}\right| x-x' \right>e^{-2iyx'/\hbar}dx'$] for the preparation of two- and five-component Schr\"{o}dinger cat state, respectively. Notice that, if we would like to dynamically generate the lower $p$-component Schr\"{o}dinger cat states ($p = 2$), the qubit-optomechanical couplings are as strong as $g = \lambda = 0.5$. In this sense, the strong-to-moderate single-photon coupling strength $g \sim 0.5$ might undermine the nonclassical optical production as it remains challenging from an experimental point of view. Furthermore, as $2plg^2 = 1$, it implies that $g$ is lower bounded by $g \geq 1/\sqrt{4l}$, and thus by reducing $g$ to $10^{-2}$ (a more experimentally available optomechanical strength) will necessarily entail letting the system to evolve for $l \sim 10^3$ oscillator's roundtrips; however, intracavity light photons for such $l$ may have leaked from the cavity at that time. One way to overcome this obstacle is to steer the light field into a nonclassical state, i.e., by projecting the qubit such as we can generate a two-component Schr\"{o}dinger cat in the weak single-photon regime. To observe this mechanism, let us write the normalized optical field projected on $\sqrt{2}\ket{+} = \up + \down$

\begin{eqnarray}
 \nonumber \hat{\rho}_c(t = 2l\pi) &=& \frac{1}{\mathcal{P}}\sum_{n,m=0}^\infty \frac{\alpha^{n+m}}{\sqrt{n!m!}}e^{2ig^2l\pi(n^2-m^2)}\\
 &\times&\cos(4\pi gl\lambda n)\cos(4\pi gl\lambda m) \ket{n}\bra{m},
\end{eqnarray}

with normalization $\mathcal{P} = \sum_{n=0}^\infty \frac{\alpha^{2n}}{n!}\cos^2(4\pi g l \lambda n)$. Moreover, we require to lighten the conditions found above for the case of  $p$-component cat state generation, as the weak regime $g \sim 10^{-2}$ is not considered in the above description. For the conditioned density matrix above, we can see that the cosine's angle should stand for $4\pi gl\lambda = \pi / 2$ to generate a two-component cat state, i.e., $g = 1/(8 l \lambda)$. The proper rate between $\lambda$ and the oscillator's cycle can give rise to $g \sim 10^{-2}$. For instance, by choosing $\lambda = 1$ and $l = 10$, makes $g \sim 0.012$ we can readily obtain nonclassical states for the optics within this regime [see Fig. \ref{fig:p-cats}-(c)]. It is also relevant to point out that, no nonclassical states due to the tripartite dynamics alone were reported with $g \sim 0.012$ and $\lambda = 1$ at times before $t = 10 \times 2\pi$ [see Fig. \ref{fig:p-cats}-(d)]; only by measuring the qubit one can collapse the optical field into a nonclassical state as shown in Fig. \ref{fig:p-cats}-(c). The hybrid quantum system, as discussed previously when $g \ll \lambda$, makes the typical nonlinear Kerr-like dynamics to evolve very slowly. In other words, the initial coherent amplitude $\alpha$ (which dynamically couples to different mechanical amplitudes due to the presence of the qubit), requires several mechanical oscillations to exhibit nonclassical fringes (i.e., negative values in the Wigner function). This is important as the Schr\"{o}dinger cat state shown in Fig. \ref{fig:p-cats}-(c) suggests that in earlier times, the ``separation'' between most probable optical coherent distributions should have been situated closer in the phase space; therefore, displaced ``kitten'' state (a Schr\"{o}dinger cat state with modest amplitude) of the optical mode can be prepared. To quantify this, we compare the actual displaced Fock number state defined in Fock basis as:

\begin{eqnarray}
 \nonumber \hat{D}(\alpha)\ket{n} &=& e^{-\frac{\alpha^2}{2}}\sum_{r=0}^{\infty} \frac{\alpha^r}{r!}\sum_{j=0}^{n}\frac{(-\alpha)^j}{j!}\sqrt{ \frac{(n-j+r)!n!}{(n-j)!(n-j)!}}\\ &\times& \ket{n-j+r},
\end{eqnarray}

with the achieved (normalized) state after qubit projection. We contrast these states using the fidelity $\mathcal{F} = |\bra{n}\hat{D}^\dagger(\alpha)\ket{\psi}_c|^2$, where $\ket{\psi}_c = \mathcal{P}^{-1/2}\sum_{n=0}^\infty \alpha^n/\sqrt{n!}e^{2ig^2l\pi n^2}\cos(4\pi gl\lambda n)\ket{n}$; $n = 1$ should be expected from a two component ``kitten'' state.

Even in a standard bipartite optomechanical system, it is known that measurements on the cavity field can conditionally project the mechanics to nonclassical states \cite{sougato}. However, in practice, a qubit (being a digital measurement of 0 or 1) may be measured much more faithfully in comparison to a continuous position degree of freedom. Thus it is relevant to find whether nonclassical states of a cavity field can be prepared even when a qubit and the field are interfaced indirectly through the mechanical element. Here we show that it is indeed the case.

\section{Concluding Remarks}\label{sec-conc}

We show how parametric coupling can induce maximal qubit-cavity entanglement in an oscillator-mediated qubit-optomechanical system. Because of the dispersive off-resonant qubit and photonic interaction, it is not readily evident that the qubit-cavity will correlate maximally at some specific time. Here, we demonstrate that the maximum value of the indirect qubit-cavity entanglement is a consequence of the complete oscillator disentanglement from the rest of the parties at each cycle. In addition to this, we show that the maximal value of the qubit-cavity negativity at the optimal time is independent of the phonon occupancy number for mechanical thermal mixtures. Thus, proving that the oscillator acts as a mediator between the qubit and the optics, where no need for cooling the mechanical oscillator to its ground state is required.

With the usage of the intrinsic entanglement $\mathcal{E}(t)$, we show that to attain such a maximum value one requires to evolve the system into the moderate-to-strong single-photon optomechanical regime, while the qubit frequencies are comparable to the mechanical oscillator frequency $\lambda \sim 1$. Because of this operational regime, we solved the master equation in the Born-Markov approximation considering an optomechanical dressed picture, which translates into a more general decoherence evolution. For feasible damping rates of the oscillator with $Q \sim 10^5$, and photon leaking rates of $\kappa \sim 10^{-2}$, we computed that the qubit can be accommodated up to $\{\Gamma, \gamma_\phi\} \sim \{10^{-3}, 10^{-2}\}$ when in contact to a thermal reservoir of $n_\mathrm{th} \sim 10$ (increasing $n_\mathrm{th} \sim 10^{2}$ requires to reduce $\Gamma$ one order of magnitude).

As a step forward, we also illustrate how the generation of nonclassical states for the cavity field can be accomplished via evolution alone, or by collapsing the cavity field through a local measurement on the qubit state ---nonclassicality evidenced by considering the Wigner quasi-probability distribution. It is known that optomechanical systems can give rise to nonclassical states of the cavity field when operating in the moderate-to-strong nonlinear single-photon regime \cite{sougato}. However, for the second case, we demonstrate that multicomponent Schr\"{o}dinger cat state and displaced one-phonon number $\hat{D}[\alpha]\ket{1}$ can also be synthesized by projecting the qubit. In practice, the advantage of a qubit projection is that a qubit may be measured much more faithfully in comparison to a continuous position degree of freedom as reported previously. Our proposal, a potential integrated hybrid node in the absence of the linearized optomechanical regime and external driving, may open up the scope for quantum networking schemes even when the interactions are not of the energy exchange type, such as purely Jaynes-Cummings type.



\end{document}